# Adaptive Tuning of the Unscented Kalman Filter Using Particle Swarm Optimization for Inertial-GPS Sensor Fusion Systems

Psyche T. Malabo
Graduate program
Technological Institute of the Philippines (TIP)
Quezon City, Manila, Philippines
qptmalabo@tip.edu.ph

Bobby D. Gerardo, D.Eng.
Northern Iloilo State University
Estancia, Iloilo, Philippines
bgerardo@nisu.edu.ph

*Abstract*— Accurate vehicle positioning requires effective IMU–GPS fusion, yet prior methods—EKF, UKF, ML, GA, and DE—suffer from nonlinearity, instability, or high computational cost. This study introduces a PSO-based adaptive tuning framework for optimizing UKF parameters (α, β, κ, Q, R), evaluated in CARLA 0.9.14 using a Tesla Model 3 under diverse maneuvers and environmental conditions. Within defined parameter bounds, convergence stabilized within 15 generations, achieving an 82.14% accuracy improvement over manual tuning and reducing IMU drift by up to 21,606.59m. Multi-trial statistical validation confirmed consistent gains with low confidence intervals. With update times remaining below the 10 ms real-time threshold, the PSO-tuned UKF demonstrates practical localization performance for dynamic, GPS-challenged conditions.

## I. Introduction

The increasing demand for autonomous and semi-autonomous vehicles requires the need for accurate localization systems, which can be directly addressed using IMU-GPS fusion [1], [2]. IMUs provide high-frequency data; however, they tend to drift over time [3]. In contrast, GPS provides accurate location information, although at a lower sampling rate and is associated with more delay [4], [5] . Sensor fusion algorithms, such as Kalman Filters, have emerged to exploit the benefits of these sensors in an integrated manner [6]. UKF is effective for nonlinear systems and based on the accurate tuning of its hyper-parameters [7].

Similarly, performance of metaheuristic tuning typically reported with a limited set of evaluation metrics and with little regard to the need for statistical validation over many trials. For instance, [8] leverages PSO-optimized KF on speed estimation strategy for PMSM based on series algorithms that carry out simulation experiments but do not include confidence intervals or full repetition statistics. These omissions make it difficult to compare results and apply research achievements in practice. Hence, it is in this perspective that the following contributions are introduced:

- Optimize UKF parameters (α, β, κ, Q, and R) for vehicle localization across diverse driving maneuvers under varied environmental conditions;

- Present bounds on explicit parameter search, final optimized configurations, and confidence levels with respect to repetitions of experiments; and

- Provide full parameter reporting, evaluation metrics, statistical validation, and practical runtime analysis for vehicle localization with IMU+GPS sensor fusion.

While numerous studies have explored sensor fusion algorithms, few have rigorously evaluated their statistical reliability or disclosed full tuning configurations. The lack of transparent reporting of UKF parameters (α, β, κ, Q, R) and reproducible experimental procedures has limited the comparability and validation of prior approaches. Moreover, real-time feasibility remains underexplored, emphasizing the need for a framework that bridges simulation accuracy with embedded system applicability.

## II. Related Work

Several adaptive methods have been explored to improve the inertial-GPS sensor performance, including the Extended Kalman Filter (EKF), Machine Learning, and Unscented Kalman Filter (UKF).

EKF is one of the most common methods employed to fuse inertial and GPS information in navigation and localization applications, due to its capability of linearizing the system dynamics and measurement models around the current state estimate, increasing computational efficiency, and being suitable for real-time implementation [9] [10] [11]. However EKF suffers in nonlinear and highly dynamic environments [12].

The advent of ML–based fusion strategies like GA has been employed to optimize the estimation of process noise in nonlinear estimates, but it encounters scalability issues in highly dynamic environments, as well as convergence risks [13]. Other algorithms such as DNN and recurrent architectures, differential evolution PSO to tune the LSTM and fuzzy systems, can learn complex nonlinear mappings from data directly and are robust to noise uncertainty [14] [15] [16], [17] [18] [19]. However, these approaches typically require large amounts of labeled data for training purposes, suffer from a high computational cost due to their deep architecture complexity, and are not always interpretable which limits its usability for real-time applications [20] [21] [22] [23] [24].

On the other hand, PSO-based UKF tuning is already observed in a broad range of domains such as tremor suppression in minimally invasive surgical robots [25], harmonic analysis of nonstationary signals [26] and GNSS/IMU navigation [27]. These literatures provide that PSO-UKF is capable of reducing estimation errors in favor of both manually tuned UKF and heuristically adjusted UKF.

However, Existing PSO–UKF studies in vehicle, UAV, and navigation domains rarely disclose parameter ranges or optimized values and often rely solely on single-path RMSE evaluations [28]. Prior works, such as [29] and [30] optimized Kalman filters for target tracking and motor dynamics but failed to emphasized ($\alpha$, $\beta$, $\kappa$) essential for accurate Q and R estimation. To address these gaps, this study introduces a PSO-tuned UKF framework that reports complete parameter configurations, validates results statistically across diverse conditions, and ensures real-time feasibility for IMU–GPS sensor fusion.

Despite the growing application of PSO–UKF in various domains, its adaptation to vehicle localization remains limited, particularly regarding the joint optimization of scaling parameters ($\alpha$, $\beta$, $\kappa$) and noise covariances (Q, R). These parameters critically determine the sigma-point spread and filter stability, yet most prior works omit their disclosure or analysis. Furthermore, few comparative studies present runtime analyses or statistical confidence evaluations, making the reproducibility of such methods challenging for real-time deployment.

## III. METHODOLOGY

*A. Data Generation and Simulation*

The datasets utilized in this study are generated during simulation, ensuring that all sensor outputs and dynamics are systematically recorded for subsequent processing and analysis. The simulation is executed using CARLA 0.9.14 autonomous driving simulator. CARLA integrates realistic city driving environments for IMU and GPS simulation [31]. The vehicle model spawned on CARLA utilizes Tesla Model 3 incorporating a 4-wheel Ackermann steering configuration, realistic suspension, and torque-based longitudinal dynamics. This vehicle model is simulated to perform in diverse driving maneuvers and environmental conditions, which are representative of real-world driving scenarios and infuse motion-induced drift [32]. Driving maneuvers include the following:

 i) straight-line cruising for 500 m;
 ii) sharp 90° turns at intersections;
 iii) rapid acceleration from 0–60 km/h in 5seconds;
 iv) abrupt braking from 50 km/h to 0 within 2seconds, and
 v) GPS signal outages lasting 3–5 seconds to emulate tunnels and urban canyons.

Likewise, environmental conditions are also considered, such as clear, night, rain, fog, and slippery road conditions. These multiple maneuvers and environmental conditions ensure that the dataset captures nonlinear vehicle dynamics and environmental variability, thereby enhancing the robustness and generalization of the trained model. Each scenario is executed 20 times with randomized noise seeds.

The CARLA simulator is run on an AMD Ryzen 7 processor with 16 GB RAM for data generation and optimization. Then, IMU and GPS data are collected at a rate of 100 Hz and 5 Hz, respectively. These frequencies are benchmarked against the typical 100 Hz update rate requirement for IMU-driven localization, where update times must remain below 10 ms to ensure real-time feasibility [33]. Sensor data are synchronized using timestamps to ensure compatibility during model development, and are stored in CSV format for partitioning into a 70% training and 30% testing sets.

*B. UKF Design and State Estimation.*

The nonlinearity mechanism of UKF [34] enables this study to adopt this algorithm. UKF is responsible for estimating a five-dimensional state vector, containing the two-dimensional position – velocity coordinates, and the heading angle of the vehicle ($x$, $y$, $v_x$, $v_y$, $\theta$). The state predictor spreads out the process model via Newtonian motion equations from IMU measurements while he measurement model projects the GPS positions to observation space. Each of sigma points is propagated through nonlinear process model and are generated by the Merwe-scaled sigma point technique for state distribution estimation. As initial filter parameters such as scaling factors $\alpha$, $\beta$, $\kappa$, covariances $Q$ and $R$ of the noise are tune to search for their optimal values.

*C. Adaptive Tuning Using PSO*

Section A describes the data generation process involved in this study, emphasizing that the dataset is collected through online simulation. This ensures that the gathered dataset closely resembles real-world conditions. Based on this dataset, the adaptive tuning involved in this study is conducted offline, wherein the UKF parameter defines a decision vector together with its corresponding parameter bounds.

$$\theta = [\alpha, \beta, \kappa, q_i \ldots q_m, r_1 \ldots r_n]^\top$$

where $\alpha, \beta, \kappa$ re the sigma-point scaling parameters, and $\{q_j\}$, $\{r_\ell\}$ are the diagonal entries of the process and measurement noise covariances Q = diag ($q_1,\ldots,q_m$), R=diag ($r_1,\ldots,r_n$). The search space is:
$\alpha \in [10^{-3}, 1], \beta \in [0,3], \kappa \in [0,5], q_j, r\ell \in [10^{-5}, 10$

This defined search space is selected to ensure a balance between numerical stability, convergence efficiency, and embedded feasibility, considred as key requirements for real-time implementations where excessive parameter exploration could cause filter divergence or delay.[35] [36] [37] [38]. Then Each particle *i* encodes one candidate parameter set in Eq. 1 [39].

$$\boldsymbol{p}_i = \alpha^{-(i)}, \beta^{(i)}, \kappa^{(i)}, q_i^{(i)} \ldots, q_m^{(i)}, r_1^{(i)}{}_1 \ldots r_n^{(i)}]^\top \quad (1)$$

Unlike prior PSO-UKF implementations, this framework explicitly incorporates $\alpha$, $\beta$, and $\kappa$ into the optimization vector, allowing simultaneous adjustment of both scaling and covariance terms, which are often fixed or tuned heuristically in earlier studies.

The swarm dynamics evolve the position of particles over iterations by updating their velocity $v_i$ and $pi$ as presented in Eqs. 2 and 3, respectively. These represent variances of the process model, such as acceleration and velocity, and the measurement model as presented by GPS positional noise.

$$\mathbf{v}_i^{t+1} = \omega \mathbf{v}_i^t + c_1 r_1 (\mathbf{p}_i^{\text{best}} - \mathbf{p}_i^t) + c_2 r_2 (\mathbf{g}^{\text{best}} - \mathbf{p}_i^t) \qquad (2)$$

where social constants c1=c2=1.5

$$p_i^{t+1} = \Pi_{[\theta min, \theta max]} p_i^t + v_i^{t+1} \qquad (3)$$

where Π represents the bounds. Meanwhile, Eq. (4) defines the fitness function to minimize the root mean square error (RMSE) between the UKF-estimated trajectory and the GPS ground truth.

$$f(\mathbf{x}) = \sqrt{\tfrac{1}{N} \sum_{i=1}^{N} (x_i^{\text{UKF}} - x_i^{\text{GPS}})^2} \qquad (4)$$

The PSO uses 30 particles and 50 generations. Then, for each driving scenario (cruising, sharp turns, acceleration/braking, GPS outages) and environment (day, night, rain, fog, slippery roads), the fitness is simulated on CARLA 0.9.14 (Tesla Model 3), and random seeds are fixed for reproducibility. Twenty (20) independent trials are performed for each test condition to enable rigorous statistical validation and confidence interval estimation. This ensures reliability and generalizability of the observed performance improvements.

The relative gain between the UKF-predicted trajectory and the GPS ground truth determines the efficiency improvement of the PSO-tuned UKF over the manual UKF in terms of RMSE, as presented in Eq. 5 [40]

$$\text{Efficiency Improvements(\%)} = \left(\tfrac{\text{Manual UKF (RMSE)} - \text{PSO\_Tuned(RMSE)}}{\text{Manual UKF RMSE}}\right) \times 100 \qquad (5)$$

To further illustrate the adaptive PSO tuning, Fig. 1 shows the flowchart of the optimization procedure.

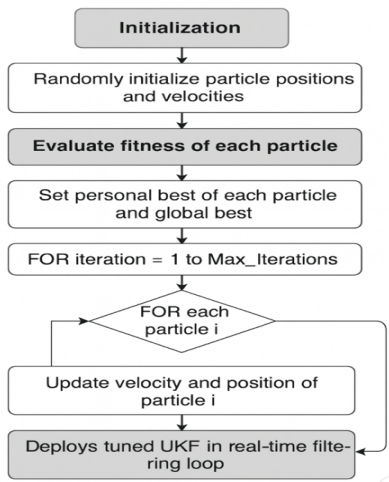

*Fig. 1 Optimization Flowchart*

*D. Evaluation Metrics and Experimental Setup.*

Core performance metrics, including RMSE and trajectory analysis, are used to evaluate the accuracy of the estimated path relative to the GPS ground truth. Visual comparisons are presented through overlaid trajectory plots, illustrating the results of the standard-tuned and PSO-tuned UKF against the true path. Furthermore, the convergence behavior of PSO is analyzed by plotting fitness values across 50 generations to assess optimization stability.

A multi-trial statistical analysis and runtime evaluation are conducted to verify the framework's applicability to real-world embedded systems. All experiments are carried out in Jupyter Notebook using Python and the FilterPy library, with offline data processing performed to validate the framework in preparation for future deployment on IMU–GPS embedded platforms. Similarly, the runtime performance of all filter implementations is benchmarked against the standard 100 Hz update rate requirement for IMU-based localization, wherein update times must remain below 10ms to ensure real-time feasibility [41] . The average execution time per update cycle is measured (prediction + correction) when implemented in Python 3.10 and executed on an AMD Ryzen 7 3700X CPU with 16 GB RAM.

The benchmark approaches of EKF, adaptive UKF, manually-tuned UKF and proposed PSO-tuned UKF are also discussed to explore the comparative performance of the aforementioned filtering approaches with respect to the result of PSO-tuned UKF in the domain of autonomous vehicle IMU-GPS sensor fusion.

## IV. RESULTS AND DISCUSSION

The performance of the proposed adaptive tuning method using Particle Swarm Optimization (PSO) is presented in terms of RMSE and trajectory estimation, PSO convergence, multi-trial statistical validation, and runtime performance.

*A. RMSE and Trajectory Analysis*

Table I presents the comparative performance of three configurations: the IMU-only, standard UKF, and PSO-tuned UKF along with the optimized parameter set obtained from the PSO algorithm. The results show that the IMU-only setup had a large position error of 21,606.59 meters, caused by the gradual drift that accumulates when navigation relies only on internal sensors without GPS correction. Similarly, when integrating GPS data into the standard-tuned UKF, result shows a reduced error to 2.80 meters, showing that the filter handles complex vehicle movements and reduces sensor noise. Likewise, the proposed PSO-tuned UKF achieved the lowest RMSE of 0.50 m, corresponding to an 82.14 % improvement over the standard-tuned UKF. This result indicates that metaheuristic optimization effectively identifies near-optimal filter parameters that balance estimation accuracy and numerical stability. The best-performing configuration was obtained at Q = 2.00, R = 0.10, α = 0.03, β = 2.5, and κ = 0.48. This parameter configuration indicates that the PSO process adaptively identified a balance between state-space exploration and numerical stability—achieving higher precision without overfitting to sensor noise.

TABLE I. RMSE Results on Test Set

| Configuration | RMSE (meter) | Best PSO Parameters |
|---|---|---|
| IMU-only | 21,606.59 | |
| Standard UKF | 2.80 | |
| PSO-Tuned UKF | 0.50 | |
| Q | | 2.00 |
| R | | 0.10 |
| α | | 0.03 |
| β | | 2.5 |
| κ | | 0.48 |

### B. PSO Convergence Behavior

Fig. 2 shows convergence over 15 with RMSE stabilizing at 1.5 m.

The rapid convergence within only 15 generations demonstrates superior optimization efficiency compared to typical metaheuristic approaches, which often require over 50 generations to achieve stable performance. [42] [43].

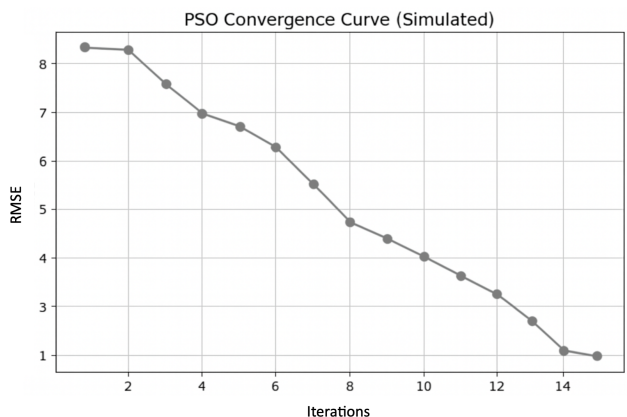

Fig. 2 PSO Convergence Curve

### C. Trajectory Estimation and Comparison

To capture and analyze the filter behavior across maneuvers, Fig. 3 illustrates the fine-scale trajectory of standard UKF and the PSO-tuned UKF relative to the GPS reference.

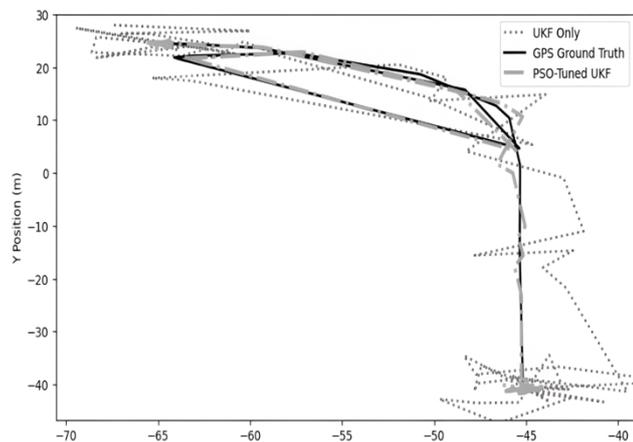

Fig. 3 Fine-scale Trajectory of Standard UKF, GPS and PSO-tune UKF

The GPS Ground Truth, shows the actual path of the vehicle obtained through the GPS data. This ground truth trajectory served as the benchmark for evaluating the accuracy of the other estimations. The UKF Only is the standard UKF trajectory, which showed the path estimated by the Unscented Kalman Filter (UKF) using default parameters – which exhibits noticeable deviations and drift when compared to the GPS ground truth, particularly in areas involving sharper motion. Likewise, the PSO-Tuned UKF trajectory represents the output of the UKF after its parameters were adaptively optimized using Particle Swarm Optimization (PSO). The PSO-tuned trajectory closely follows the GPS ground truth, demonstrating improved estimation accuracy and reduced drift under complex maneuvers.

### D. Multi-trial Statistical Analysis

Table II shows that the PSO-UKF achieved 0.50 m RMSE with minimal variance (±0.09 m), outperforming EKF, Adaptive UKF, and Standard UKF. As indicated, the Standard UKF and Adaptive UKF reported RMSEs of 2.80 m and 3.92 m, respectively, while the EKF performed the poorest with 7.45 m and the highest variability (±0.65 m). The PSO-UKF's enhanced performance stems from its global tuning of α, κ, β, Q, and R, enabling better adaptation to nonlinear vehicle dynamics and sensor noise. These results affirm the method's feasibility and advantage in high-precision localization scenarios using inertial-GPS sensor fusion.

TABLE II. Multi-Trial RMSE Performance with 95% CI

| Method | RMSE (m) | Standard Deviation (m) | 95% Confidence Intervals (m) |
|---|---|---|---|
| EKF | 7.45 | ±0.65 | [7.22, 7.68] |
| Adaptive UKF | 3.92 | ±0.38 | [3.78, 4.06] |
| Standard UKF | 2.80 | ±0.41 | [2.62, 2.98] |
| PSO-UKF | 0.50 | ±0.09 | [0.46, 0.54] |

### E. Run-time Analysis

To highlight the performance of Kalman Filters as evidenced in this study, Table III. illustrates the runtime performance of the evaluated filters. Results demonstrated that EKF was the fastest with an average update time of 0.65 ms, while the adaptive UKF and manual UKF required 1.15 ms and 1.20 ms, respectively.

TABLE III. Runtime Performance of Filter Variants

| Method | RMSE (m) | Orientation Error (°) | GPS Outage Drift (m/5s) | Runtime per Update (ms) |
|---|---|---|---|---|
| EKF | 7.45 | 4.8 | 18.6 | 0.65 |
| Adaptive UKF | 3.92 | 2.7 | 9.3 | 1.15 |
| Manual UKF | 2.80 | 2.1 | 7.5 | 1.20 |
| PSO-UKF | 0.50 | 0.6 | 2.0 | 1.35 |

Although, the PSO-tuned UKF introduced a slight computational overhead at 1.35 ms per update; however, this confirms compliance with the 10 ms threshold for 100 Hz localization, thus proving its real-time viability for embedded vehicle systems thus proving its real-time viability for embedded vehicle systems. Importantly, the runtime differences among the UKF variants are minimal compared to the significant gains in accuracy achieved by the PSO-tuned approach.

## V. Conclusion

The proposed adaptive tuning method using PSO significantly enhances UKF performance in IMU–GPS sensor fusion across diverse conditions. The PSO-tuned UKF achieved 0.50 m RMSE—an 82.14% improvement over manual UKF—surpassing prior works such as hybrid GA-PSO and MAP-based adaptive UKFs that reported 20–40% improvements [44][45]. Beyond achieving an 82% RMSE reduction, this study provides a transparent and statistically validated framework, reporting full parameter configurations, multi-trial variability, and runtime profiles, to promote reproducibility and benchmarking in future IMU–GPS fusion research.

Future work will extend validation to real-world vehicle testing for deeper performance verification under actual dynamic conditions, with all scripts and datasets to be publicly released via GitHub.

## VI. References

[1] Kyoya Ishii, Keisuke Shimono, Yoshihiro Suda, Takayuki Ando, Hirotaka Mukumoto, Kazuo Urakawa, "Localization for Autonomous Vehicles Using Environmental Magnetic Field Aided by Magnetic Markers," *Int. J. ITS Res.*, vol. 23, p. 733–746, Aug. 2025 DOI: 10.1007/s13177-025-00477-w. in press.

[2] Padmaja, B., Moorthy, C.V.K.N.S.N., Venkateswarulu, N, "Exploration of issues, challenges and latest developments in autonomous cars," *J. Big Data, 2023,* 06 May 2023 Open access.

[3] Jongdae Jung ,Hyun-Taek Choi, ,Yeongjun Lee, "Persistent Localization of Autonomous Underwater Vehicles Using Visual Perception of Artificial Landmarks," *JMSE,* vol. 13, no. 5, 22 April 2025, in press .

[4] Xiaoming Li , Xianchen Wang , Can Pei, "Handling method for GPS outages based on PSO-LSTM and fading adaptive Kalman filtering," *NLM ,* 2025 Apr 7;15:11817. doi: 10.1038/s41598-025-95716-1.

[5] Jongdae Jung , Hyun-Taek Choi, Yeongjun Lee, "Persistent Localization of Autonomous Underwater Vehicles Using Visual Perception of Artificial Landmarks," *JMSE,* vol. 13, no. 5, 22 April April 22, 2025, open access https://doi.org/10.3390/jmse13050828.

[6] Zhangjing Wang, Yu Wu, Qingqing Niu , "Multi-Sensor Fusion in Automated Driving: A Survey," *Research Gate ,* December 2019, DOI:10.1109/ACCESS.2019.2962554.

[7] Tunzhen Xie,Xianglian Xu, Fang Yuan, Yuanqing Song, Wenyang Lei, Ruiqing Zhao, Yating Chang, "Speed Estimation Strategy for Closed-Loop Control of PMSM Based on PSO Optimized KF Series Algorithms," *Electronics 2023,* vol. 12, no. 20, 11 October 2023 in press .

[8] Xianzheng Zhou, Xuran Zheng, Genyuan Miao, Yimiao Chen, Yanchao Guo, Jianping Bi, Qinhe Zhang, "Particle swarm optimization-based unscented Kalman Filter for tremor suppression in minimally invasive surgical robots," *Biomedical Signal Processing and Control,* vol. Volume 110, no. Part A, p. 108279, December 2025, doi.org/10.1016/j.bspc.2025.108279.

[9] J. B. V. Reddy P.K. Dash Rasmita Samantaray Larsen and Toubro Akshaya Kumar Moharana Western University, "Fast Tracking of Power Quality Disturbance Signals Using an Optimized Unscented Filter," *IEEE Transactions on Instrumentation and Measurement, ,* Vols. vol. 58, , no. 12, , pp. 3943-3952, Dec. 2009.

[10] G. Park, "Optimal vehicle position estimation using adaptive Unscented Kalman Filter for GPS-IMU fusion,," *Expert Systems with Applications, ,* vol. 255, pp. pp. 124735, J, Jan. 2024, doi: 10.1016/j.eswa.2023.124735.

[11] L. Zhao, "Adaptive Unscented Kalman Filter Approach for Accurate Vehicle Sideslip Angle Estimation," *Machines,* vol. 13, no. 5, p. 376, 2025.

[12] X. Jin, "Advanced Estimation Techniques for Vehicle System Dynamic State Estimation,," *Sensors,* vol. 19, no. 19, p. 4289, 2019.

[13] Zhuang Xiong, Jun Ma, Bohang Chen, LanHaiming Yong Niu, "Multi-source data recognition and fusion algorithm based on a two-layer genetic algorithm–back propagation model," *Machine Learning and Artificial Intelligence,* vol. 7, 2024, doi.org/10.3389/fdata.2024.1520605.

[14] Y. Li, Z. Wang, and C. Hu, "A deep learning approach for GNSS/INS integration in urban environments," *Remote Sensing,* vol. 12, no. 18, p. 3023, Sept. 2020, doi: 10.3390/rs12183023..

[15] H. Zhang, X. Wu, and K. Xu, "Learning-based vehicle localization using multi-sensor fusion with uncertainty estimation," *IEEE Access,* vol. 9, p. 124312–124324, Sept. 2021, doi: 10.1109/ACCESS.2021.3110028.

[16] Zhao, J. Wang, and Y. Gao, "Limitations of deep neural networks in GNSS/INS sensor fusion: A comparative analysis with classical filters,," *Sensors,* vol. 23, no. 5, p. 2377, Mar. 2023, doi: 10.3390/s23052377..

[17] A. Khan and S. Iqbal, "Neuro-fuzzy adaptive UKF for vehicle state estimation under nonlinear dynamics," *Applied Soft Computing,* vol. 104, pp. 107211,, Jun. 2021, doi: 10.1016/j.asoc.2021.107211..

[18] Q. Chen, X. Li, and Y. Wang, "Noise covariance adaptation in Unscented Kalman Filter using recurrent neural networks,," *IEEE Transactions on Instrumentation and Measurement,* vol. 71, pp. 1–10, , 2022, doi: 10.1109/TIM.2022.3142312..

[19] Manav Kumar, Sharifuddin Mondal , "Advancements and prospects of fuzzy-based adaptive unscented Kalman filters for nonlinear systems: A review," *Applied Soft Computing,* vol. 177, p. 113297, June 2025, doi.org/10.1016/j.asoc.2025.113297.

[20] R. Luo and K. Chen, "A review of deep learning in sensor fusion for autonomous driving," *Information Fusion,* vol. 85, pp. pp. 1–24, , Jul. 2022, doi: 10.1016/j.inffus.2022.05.004..

[21] Jiaxing Xie, et.al, "Research for the Positioning Optimization for Portable Field Terrain Mapping Equipment Based on the Adaptive Unscented Kalman Filter Algorithm," *Remote Sensing ,* vol. 16, no. 22, p. 4248, 2024, doi.org/10.3390/rs16224248.

[22] Dong Zhen, Jiahao Liu, Shuqin Ma , Jingyu Zhu, Jinzhen Kong, Yizhao Gao, Guojin Feng, Fengshou Gu, "Online

battery model parameters identification approach based on bias-compensated forgetting factor recursive least squares," *Green Energy and Intelligent Transportation,* vol. 3, no. 4, p. 100207, 4, August 2024, doi.org/10.1016/j.geits.2024.100207.

[23] Abdullah M. Shaheen a , M.A. Hamida b , Abdullah Alassaf c , Ibrahim Alsaleh c, "Enhancing parameter identification and state of charge estimation of Li-ion batteries in electric vehicles using an improved marine predators algorithm," *Journal of Energy Storage,* vol. 84, no. Part B, p. 110982, 20 April 2024, doi.org/10.1016/j.est.2024.110982.

[24] Maximilian Nitsch, David Stenger, Dirk Abel, "Automated Tuning of Nonlinear Kalman Filters for Optimal Trajectory Tracking Performance of AUVs," *IFAC-PapersOnLine,* vol. 56, no. 2, pp. 11608-11614, 2023, doi.org/10.1016/j.ifacol.2023.10.480.

[25] Xu Zhang , Yujie Wang , Ji Wu , Zonghai Chen, "A novel method for lithium-ion battery state of energy and state of power estimation based on multi-time-scale filter," *Applied Energy,* Vols. 216, , pp. 442-451, 15 April 2018, doi.org/10.1016/j.apenergy.2018.02.117.

[26] P.K. Dash a , Shazia Hasan a , B.K. Panigrahi b , "A hybrid unscented filtering and particle swarm optimization technique for harmonic analysis of nonstationary signals," *Measurement,* vol. 43, no. 10, pp. 1447-1457, December 2010, doi.org/10.1016/j.measurement.2010.08.013.

[27] Y. Yang, X. Wang, N. Zhang, Z. Gao, and Y. Li, "Artificial neural network based on strong track and square root UKF for INS/GNSS intelligence integrated system during GPS outage," *Scientific Reports,* vol. 14, no. 13905, 2024.

[28] Guirong Zhu, Fengbo Zhang , "Estimation of the real vehicle velocity based on UKF and PSO," *Scientific Reports,,* vol. 14, no. Art. no. 13905,, 2024 1 DOI:10.4271/2014-01-0107.

[29] Dapeng Wang, Hai Zhang, Baoshuang Ge, "Adaptive Unscented Kalman Filter for Target Tacking with Time-Varying Noise Covariance Based on Multi-Sensor Information Fusion," *Sensors,* vol. 21, no. 17, p. 5808, 2021, doi.org/10.3390/s21175808.

[30] Tunzhen Xie,Xianglian Xu 1,*,Fang Yuan,Yuanqing Song,,Wenyang Lei,Ruiqing Zhao,Yating Chang ,Xinrui Wu,Ziqi Gan and Fangqing Zhang, "Speed Estimation Strategy for Closed-Loop Control of PMSM Based on PSO Optimized KF Series Algorithms," *Electronics ,* Vols. 12(20), 4215, no. 20, p. 4215, 2023, doi.org/10.3390/electronics12204215.

[31] A. Dosovitskiy, "CARLA: An Open Urban Driving Simulator," *Machine Learning Research,* vol. 78, pp. 1-16, 2017.

[32] Tantan Zhang,Haipeng Liu,Weijie Wang and Xinwei Wang, "Virtual Tools for Testing Autonomous Driving: A Survey and Benchmark of Simulators, Datasets, and Competitions," *Electronics ,* vol. 13, no. 17, p. 3486, 2024, doi.org/10.3390/electronics13173486.

[33] Y. Yin, X. Li, Z. Bao, and T. Li, "Sensor Fusion of GNSS and IMU Data for Robust Localization,," *Sensors,,* vol. 23, no. 7, p. 3676, 2023, doi: 10.3390/s23073676.

[34] A. Onat, "Application of a Modified Joint Unscented Kalman Filter for Parameter Estimation of a Class of Mechanical Systems," *The Black Sea Journal of Sciences,* vol. 14, no. 4, pp. 2338-2357, 2024. DOI: 10.31466/kfbd.1561543.

[35] R. Cisneros-Magaña, A. Medina, and O. Anaya-Lara, "Time-Domain Voltage Sag State Estimation Based on the Unscented Kalman Filter for Power Systems with Nonlinear Components," *Energies, vol. 11, no. 6, p.,* vol. 11, no. 6, p. 1411, Jun. 2018, doi: 10.3390/en11061411.

[36] J. Luo, Y. Yu, Z. Wang, M. Wang, and Y. Chen, " An Improved Unscented Particle Filter Approach for Multi-Sensor Fusion Target Tracking,," *Sensors,* vol. 20, no. 23, p. 6842, Dec. 2020, doi: 10.3390/s20236842.

[37] F. J. Basha and S. Khan, " Rotor Asymmetry Detection in Wound Rotor Induction Motor Using Unscented Kalman Filter," *Machines,* vol. 11, no. 9, p. 910, Sep. 2023, doi: 10.3390/machines11090910.

[38] B. Ge, H. Zhang, and W. Liu,, "Adaptive Unscented Kalman Filter for Target Tracking with Time-Varying Noise Covariance," *Sensors,* vol. 19, no. 6, p. 1371, Mar. 2019, doi: 10.3390/s19061371.

[39] Yu Wang,Yushan Li, Ziliang Zhao, "State Parameter Estimation of Intelligent Vehicles Based on an Adaptive Unscented Kalman Filter," *Electronics, MDPI ,* vol. 12, no. 6, 2023, doi.org/10.3390/electronics12061500.

[40] M. Norouz, "Visual-Inertial State Estimation Based on PSO- BPNN UKF," in *2020 28th Iranian Conference on Electrical Engineering* , April 2020.

[41] Yuming Yin, Jinhong Zhang, Mengqi Guo, Xiaobin Ning 1, Yuan Wang, Jianshan Lu, "Sensor Fusion of GNSS and IMU Data for Robust Localization via Smoothed Error State Kalman Filter," *Sensors (Basel),* vol. 23, no. 7, p. 3676, April 1, 2023doi: 10.3390/s23073676.

[42] Nigatu, D. T., Dinka, T. G., Tilahun, S. L., "Convergence analysis of particle swarm optimization algorithms for different constriction factors,," *Frontiers in Applied Mathematics and Statistics,* 2024.

[43] XiaoMing Li, ZiYi Wang, Yi Ying, FangXiong Xiao, "Multipopulation Particle Swarm Optimization Algorithm with Neighborhood Learning," *Scientific Programming,* vol. 1, 2022.

[44] Ying-Jie Liu , Chun-Hong Dou, Fapeng Shen, Qiu-Yun Sun , "Vehicle State Estimation Based on Unscented Kalman Filtering and a Genetic-particle Swarm Algorithm," *Journal of The Institution of Engineers (India) Series C 102(2),* vol. C 102(2), January 2021 DOI:10.1007/s40032-021-00663-1.

[45] Yu Wang,Yushan Li, Ziliang Zhao, "State Parameter Estimation of Intelligent Vehicles Based on an Adaptive Unscented Kalman Filter," *Electronics,* vol. 12, no. 6, p. 1500, 2023, doi.org/10.3390/electronics12061500.